\documentclass[%
amsmath,
aps, 
prl,
twocolumn,
]{revtex4-2}

\usepackage{graphicx}
\usepackage{dcolumn}
\usepackage{bm}
\usepackage[T1]{fontenc}
\usepackage[utf8]{inputenc}
\usepackage{hyperref}
\usepackage{placeins}
\usepackage{ulem}
\usepackage{circuitikz} 
\usepackage{tikz}
\usetikzlibrary{patterns,decorations.pathmorphing,decorations.markings,shapes,arrows,positioning,decorations.pathreplacing,calc,backgrounds,external,spy}
\usepackage{float}

\begin{document}

\title{\textbf{Transit-time oscillations in nanoscale vacuum diode with a pure resistive load} 
}

\author{Bjart\thór Steinn Alexandersson}
\author{Kristinn Torfason}%
\author{Andrei Manolescu}%
\author{Ágúst Valfells}%
 \email{Contact author: av@ru.is}
\affiliation{Department of Engineering, Reykjavik University, Menntavegur 1, IS-102, Reykjavik, Iceland.
}


\begin{abstract}
We examine the Ramo current in a nanoscale planar vacuum diode undergoing field emission in the presence of a DC voltage supply and an external resistor. We describe a simple mechanism for generating persistent current oscillations in the diode due to the voltage drop across the external resistor (beam loading) which reduces the total field and inhibits the emission.  The amplitude and the frequency, which is in the THz domain, depend on the operating parameters of the diode. Molecular dynamics simulations are used to find the characteristics and physical basis of the mechanism, and a simple analytical model is presented, in good agreement with the simulation. 
\end{abstract}

\maketitle

Oscillations in vacuum diodes have been studied for many decades \cite{Benham_1928, Muller_1934, Llewellyn_1939}.  Initially in the MHz domain, the frequency of such oscillations later reached GHz and even THz with technological advancements and device miniaturization. In general, transit-time-based oscillations are used to amplify the high-frequency signal of a basic RLC circuit, such as an electromagnetic cavity, when the motion of electrons is at resonance with the frequency of that circuit \cite{Kwan_1991,Barroso_1999}.  A higher amplification was achieved by imposing oscillations of the electron beam transverse to the direction of the net current through the diode, in coaxial structures \cite{Mostrom_1995,Arman_1996}. Usually the space charge associated with the electron distribution inside the diode gap plays an important role and theoretical papers considered that using the Particle-in-Cell (PIC) computational method \cite{Luginsland_1997, Benedik_2013}.

Examples where the diode itself is a source of oscillations, without the requirement of a resonant RLC circuit, are rare. Space-charge self-oscillations due to Coulomb blocking, in the THz range, have been predicted by simulations, with a photoemission mechanism, but have not yet been experimentally confirmed \cite{Pedersen_2010,Ilkov_2015}. 
Another proposed mechanism, again based on space charge, has been associated with the oscillations of the virtual cathode in a so-called nanovircator \cite{Frolov_2015,Kurkin_2016}.
More recently, intrinsic and even faster oscillations have also been predicted by simulations, based on the interplay between the field emission current and space charge in a microplasma, where the ionized atoms also play a fundamental role \cite{Levko_2020,Chen_2024}.  

For the generation of current in a vacuum diode, field emission is often the desired  emission mechanism. It is a quantum mechanical tunneling process that is enhanced by the presence of a strong electric field at the surface of the emitter.
In general, field emission is important in modern vacuum electronics due to its excellent characteristics, e.g. responsiveness to an applied electric field (for rapid switching) and low power consumption \cite{Dwivedi_2021,Zhang_2021,Zhang_2016,Li_2025,Sapkota_2021}. A challenge associated with operating field-emitter devices is the very strong dependence of the current on the surface electric field. This has implications on the stability and longevity of devices and may also decrease the uniformity of emission in arrays of emitters. One solution to this strong dependence is to apply a resistor to the field-emitting surface to flatten the characteristic current-voltage curve ~\cite{Lin_2022,Luginsland_1996}. 

In this work, we present computer simulations of the effects of such an external resistor using a molecular-dynamics code that includes self-consistent emission modeling, full Coulomb interaction between individual electrons, and calculations of induced current from electron motion in the vacuum region. In particular, we report on a simple mechanism that can lead to persistent self-oscillations in the diode current, at fixed frequency and amplitude, which do not require an external resonator. To the best of our knowledge this situation has not been captured in previous work on the effects of an external resistor. We show that these oscillations are due to the effect of beam loading on the current source, while the space-charge distribution has a negligible role.

The simulations were run with our grid-free, three dimensional molecular dynamics code RUMDEED \cite{Torfason_2015, Torfason_2021, Sitek_2021a, Sitek_2021b}, which calculates direct mutual interaction between all electrons in the vacuum gap and all their image charges. We use the self-consistent field emission module, that takes into account discrete space-charge effects from electrons in the diode gap on the local field of the cathode surface. The field emission is implemented with the Fowler-Nordheim equation~\cite{Fowler_1928}
\begin{equation}
 J=\frac{A}{t^2\left(\ell\right)\phi}E^2e^{-\nu\left(\ell\right)B\phi^\frac{3}{2}/E} \ ,
\label{eq:FowlerNordheim}
\end{equation}
where $J$~[$Am^{-2}$] is the current density, $\phi$~[eV] is the work function and $E$~[$Vm^{-1}$] is the electric field at the surface of the cathode (taken to be positive in value). The first and second Fowler-Nordheim constants are 
$A=1.54\times 10^{-6}$~[A eV V$^{-2}$] and $B=6.83\times 10^9$~[eV$^\frac{-3}{2}$ V m$^{-1}$], respectively, and $\nu\left(\ell\right)$ is the Nordheim function. We use the approximations provided by Forbes and Deane \cite{Forbes_2007} 
\begin{equation}
 \nu\left(\ell\right)=1-\ell+\frac{1}{6}\ell\ln\left(\ell\right) 
\label{eq:NordheimFunction}
\end{equation}
and
\begin{equation}
 t\left(\ell\right)=1+\ell\left(\frac{1}{9}-\frac{1}{18}\ln\left(\ell\right)\right) \ ,
\label{eq:Forbes}
\end{equation}
with $\ell=\frac{q}{4\pi\varepsilon_0}\frac{E}{\phi^2}$, where
the charge of the electron, electron mass, and permittivity of vacuum are denoted by $q$, $m$, and $\varepsilon_0$ respectively.

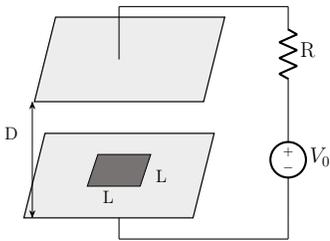
\begin{figure}
    \centering
    \resizebox{0.25\textwidth}{!}{%
    \begin{circuitikz}
    \tikzstyle{every node}=[font=\LARGE]
    \draw [ line width=0.7pt](12.25,13.5) to[R,l={ \Large R}] (12.25,11.25);
    \draw [ fill={rgb,255:red,237; green,237; blue,237} , line width=0.8pt ] (6,8.5) -- (10,8.5) -- (10.5,10.5) -- (6.5,10.5) -- cycle;
    \draw [ fill={rgb,255:red,237; green,237; blue,237} , line width=0.8pt ] (6.25,11.25) -- (10.25,11.25) -- (10.75,13.25) -- (6.75,13.25) -- cycle;
    \draw [ fill={rgb,255:red,120; green,115; blue,115} ] (7.5,9.25) -- (8.75,9.25) -- (9,10) -- (7.75,10) -- cycle;
    \node [font=\large] at (9.25,9.5) {L};
    \node [font=\large] at (8,9) {L};
    \draw [ line width=0.7pt](12.25,11.25) to[american voltage source,l={ \Large $V_0$}] (12.25,8.5);
    \draw [ line width=0.7pt](12.25,8.5) to[short] (12.25,8);
    \draw [ line width=0.7pt](12.25,8) to[short] (8.25,8);
    \draw [ line width=0.7pt](8.25,8) to[short] (8.25,8.5);
    \draw (12.25,13.5) to[short] (12.25,13.5);
    \draw [ line width=0.7pt](12.25,13.5) to[short] (8.25,13.5);
    \draw [ line width=0.7pt](8.25,13.5) to[short] (8.25,12.25);
    \draw [<->, >=Stealth] (6.2,8.5) -- (6.2,11.25);
    \node [font=\large] at (5.7,10.5) {D};
    \end{circuitikz}
    }%
    \caption{System schematic. Note that the planar diode is assumed to be infinite in extent, but the emitter area is finite.}
    \label{fig:diagram}
\end{figure}

In the following, we present simulation results that illustrate the nature of the oscillations, as well as a simple analytical model for comparison.
Our simulations are based on the physical configuration shown in Figure~\ref{fig:diagram}. It consists of an infinite-area parallel plate diode with gap spacing, $D$, applied potential, $V_0$, and external resistance, $R$. The applied potential increases linearly from zero to $V_0{_m}$ over a rise time, $t_r$, after which it remains constant. The grounded cathode may only emit from a square area of side length $L$ embedded in the plane. The emitting area has a work function of $\phi$. The time step of the simulations is 0.1 fs.

The current, $I$, is calculated using the Ramo-Shockley theorem~\cite{doi:10.1063/1.1710367, 1686997},
\begin{equation}
 I=\frac{q}{D}\sum^{N_e}_{k=1}v_{k,z} \ ,%
\label{eq:Ramo}
\end{equation}
where $v_{k.z}$ is the velocity of the $k$th electron in the $z$ direction and $N_e$ is the number of electrons in the vacuum region of the diode. The anode potential will be denoted by $V_a$ and varies with the current. The model is electrostatic, with time-varying properties. For given values of the parameters $D$, $L$, $\phi$, and $V_0{_m}$ and setting $R$ equal to zero, one obtains the current $I_0{_m}$. From this we may calculate an inherent impedance, $Z_0 = V_0{_m} / I_0{_m}$, for the diode. In our simulations, we measure the number of electrons emitted from the cathode per time step, $n_e$, and the number of electrons in the gap at each time, as well as the current in the diode, and the anode potential.

\begin{figure}[h]
  \centering
  \includegraphics[width=0.9\linewidth]{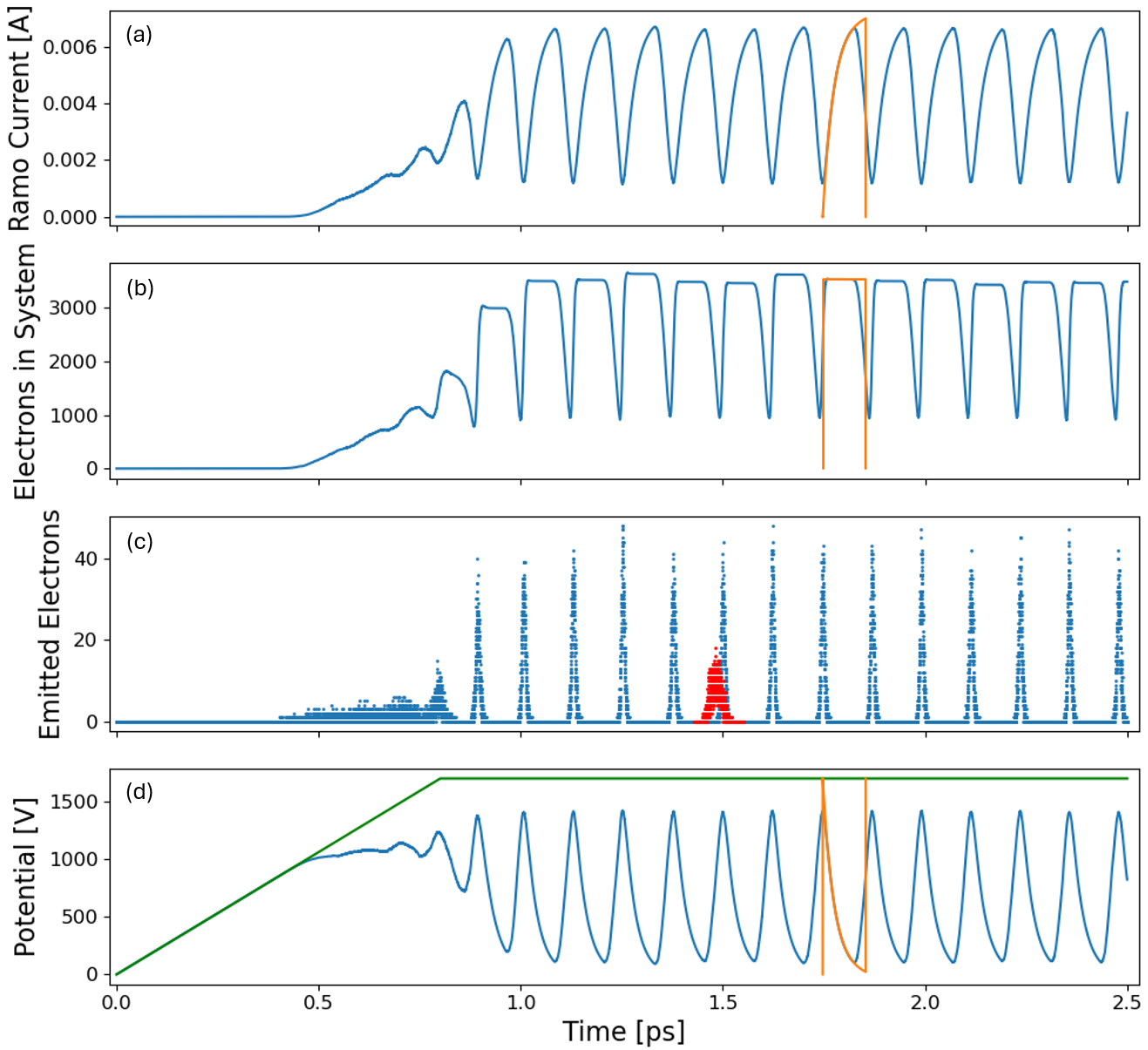}
  \caption{Evolution of the system with time. (a): Diode current. (b): Electrons in the vacuum gap. (c) The number of electrons emitted per time step, with red dots indicating electrons absorbed that originated from the immediately preceding emission event. (d): Applied potential (green) and anode potential (blue). $V_0{_m}$ = 1700 V, $D$ = 800 nm, $L$ = 1000 nm, $R$ = 240 k$\Omega$, $\phi$ = 2.0 eV, and $t_r$ = 0.8 ps. The orange curves show variations for a single period as obtained from the analytical model associated with Equations~\ref{eq:Diffeq} and~\ref{eq:position}.}
  \label{fig:VoltageIncrease}
\end{figure}

Let us now look at the simulation results using the base parameters of $V_0{_m}$ = 1.7 kV, $D$ = 800 nm, $L$ = 1000 nm, $\phi$ = 2.0 eV, $t_r$ = 0.8 ps and $R$ = 240 k$\Omega$. Figure~\ref{fig:VoltageIncrease} shows the current, the number of electrons in the gap, the number of electrons emitted from the cathode per time step, and the anode potential as a function of time. The mechanism is as follows: Initially, there is no emission because the applied potential is too low to cause field emission. As the applied potential rises, field emission is enhanced. The electrons in the gap are accelerated toward the anode, and thus the diode current rises. With increasing current comes an increasing potential drop across the external resistor (i.e., beam loading of the diode) which, in turn, leads to the anode potential $V_a$ being smaller than the applied potential $V_0$. Interestingly, we see an instability develop due to the interplay between anode voltage, emission and loading of the circuit by the diode current. Once the applied voltage has reached a stable value, we see a persistent pattern emerge. When the current is low, the anode voltage is high, and electrons are emitted from the cathode. As these electrons gain in velocity, the current in the diode grows, causing beam loading which, in turn, leads to a drop in the electric field at the cathode surface and drastically reduces emission to a vanishing level. As the cloud of electrons is accelerated across the diode gap, the current increases, and emission remains negligible. Once electrons begin to reach the anode and are removed from the gap, the current once again drops and the anode potential rises. As the surface electric field increases to a sufficient strength, field emission commences again. This pattern is repeated.

Several questions naturally arise in conjunction with these results: Is this a space-charge effect independent of the external resistor? How is the response related to the magnitude of the external resistance? Is the oscillation dependent on the rise time of the applied potential (e.g. some sort of shock effect)? What are the effects of diode parameters such as the gap spacing, emitter area, applied potential, and work function? The answers to some of these questions follow. 

We begin by noting that, in the absence of the external resistor, there is generally no oscillation. An exception to this is that there may be a slight oscillation due to space-charge effects when the rise time of the applied voltage is small compared to the time that it takes for the electrons to cross the diode gap ~\cite{Feng_2006,Torfason_2015,Jensen_2015}. This oscillation is rapidly damped. It is important to observe that, in the presence of a suitably large external resistor, the onset of the oscillation is not due to a discontinuity in the applied potential (e.g.\ an unphysically short value of $t_r$). Rather, it can grow from a small perturbation before settling into a persistent pattern. The effect of increasing external resistance is further illustrated in Supplemental Material, Figure S1. 

\begin{figure}[h]
  \centering
  \includegraphics[width=0.9\linewidth]{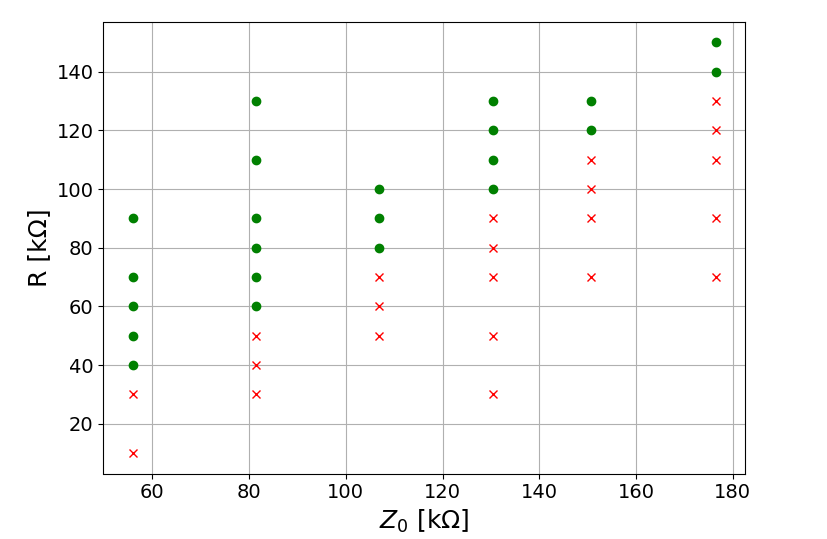}
  \caption{Map of existence of oscillation in $R - Z_0$ space. Green dots signify occurrence of persistent oscillations. Red crosses signify no persistent oscillations. $D$ = 1000~nm, $L$ = 1000~nm, $\phi$ = 2.0~eV. $R$ and $V_0{_m}$ are varied.}
  \label{fig:Rzmap}
\end{figure}

As can be seen in Figure~\ref{fig:Rzmap}, the persistent oscillations arise only when the external resistance is of the same order of magnitude as the inherent impedance of the diode at a fixed applied potential. Or, put in another way, when the potential drop across the resistor is an appreciable fraction of the applied potential. A boundary between oscillatory and nonoscillatory behavior is readily apparent. The data presented in Figure~\ref{fig:Rzmap} is based on fixed values of $D$, $L$, and $\phi$. Only the external resistance and maximum value of the applied voltage, $V_0{_m}$, are varied. Nonetheless, we find that the conditions for the onset of a persistent oscillation are not simply due to the relation between the external resistance and implicit impedance. For instance, we may consider two configurations of the diode circuit that have the same values of $D$ and $V_0{_m}$, but different values for emitter area and work function. 
The combination of parameters is chosen such that the two diodes draw the same current for the same applied voltage in the absence of an external resistor. This means they both have the same value of $Z_0$. Configuration A uses $\phi$ = 1.6 eV and $L$ = 395 nm, while configuration B uses $\phi$ = 2.0 eV and $L$ = 1000 nm. Both use the values $V_0{_m}$ = 1700 V, $D$ = 1000 nm, and $t_r$ = 0 s. 

\begin{figure}[h]
  \centering
  \includegraphics[width=0.95\linewidth]{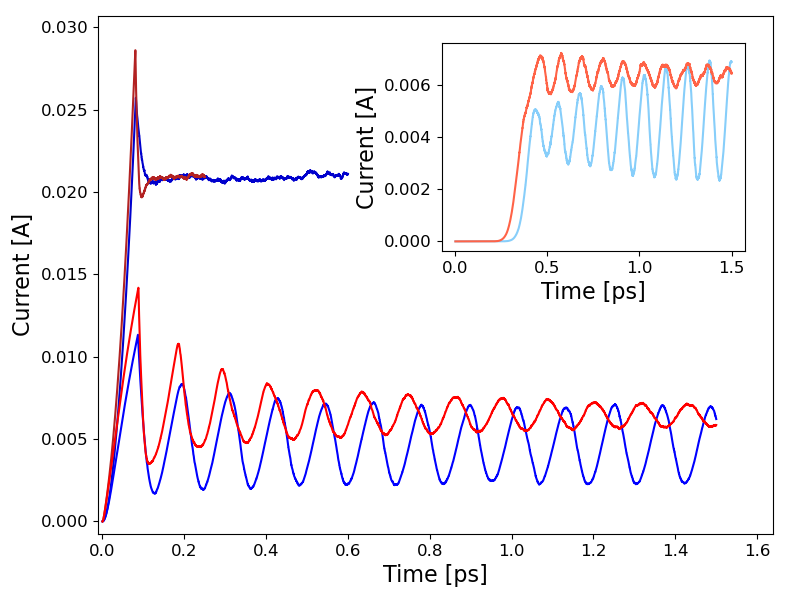}
 \caption{Comparison of two different configurations with the same $Z_0$. A Configuration A (red) has $\phi$ = 1.6 eV and $L$ = 395 nm. Configuration B (blue) has $\phi$ = 2.0 eV and $L$ = 1000 nm. Both configurations are for $V_0{_m}$ = 1700 V and $D$ = 1000 nm. Main graph shows current with (lower curves) and without (upper curves) 60 k$\Omega$ resistor and vanishing rise time for applied potential. Inset shows current for 60 k$\Omega$ resistor and 0.4 ps rise time.}
  \label{fig:longtime}
\end{figure}

A comparison of these two configurations can be seen in Figure~\ref{fig:longtime}. The figure depicts the current as a function of time. The inset does so as well, but for a finite value of the rise time for the applied potential, $t_r$ = 0.4 ps.  First, we note that in the absence of an external resistor we obtain the same steady state current for both cases. We see an initial overshoot in the current followed by a damped oscillation before settling in to the steady state. 
The explanation is that the applied potential is at maximum value immediately and, as the diode gap is being filled during the first transit, space-charge effects have not fully taken hold (this overshoot and oscillation decrease significantly if the rise time is longer than the transit of electrons across the gap). This oscillation is due to space-charge effects
and is observed to be 
more prominent for the larger initial current. This last observation agrees with earlier work. In the presence of an external resistor we see a marked change in the current response. Figure S2 of Supplemental Material shows additional data on the effects of varying the emitter area. 

For configuration A, with the lower work function, the oscillation is damped and the average current is higher than for configuration B. In configuration B the oscillation is persistent. The inset picture shows the current response for the different configurations when the potential rise time is longer than the time of transit. Even though there is an initial perturbation in the current for the low work function configuration, it dies out, in contrast to the persistent oscillation for the higher work function. The effect of the work function on current instability is related to the highly nonlinear characteristics of field emission. Our simulations reveal that a characteristic of the most distinctive oscillations is that they occur when electrons are emitted into the diode gap in short bursts. In case of a higher work function, the diode current is more strongly affected by beam loading than for the lower work function, as a drop in the anode potential leads to a significantly greater decrease of emission at the cathode. Hence, the bunching effect becomes more pronounced for a high work function than a low work function.

One key aspect is that the charge emission rate given by Eq.\ (\ref{eq:FowlerNordheim}) depends on the total field present inside the diode, $E=(V_0-IR)/D$. Another key aspect is that the Ramo-Shockley current $I$, Eq. (\ref{eq:Ramo}), depends not only on the total charge residing in the diode gap, but also on the velocities of the charge carriers, which depend on time.  Therefore, while the electrons are emitted and accelerated, the Ramo current dramatically increases, and the total field drops or even vanishes before many or even all emitted electrons have reached the anode, i.\ e. within a time interval shorter than the typical transit time. The emission process is thus reduced by the current, the current itself decreases, but then the emission increases back. This mechanism resembles a relay, where the increase of the excitation current causes the interruption of the circuit. Previous work on a similar setup did not capture this instability as it was based on a lumped circuit model with implicit DC current~\cite{Luginsland_1996}.

We present a simple analytical model for the beam loading mechanism. The model is the same as shown in Figure~\ref{fig:diagram}, except that we imagine that all the emitted charge manifests as a single sheet of charge density $\sigma = \frac{Q}{L^2}$ where $Q$ is the total charge in the diode gap. Essentially, this means that we have uniform electric fields, on either side of the sheet, accelerating it across the gap. It is straightforward to derive an equation of motion in terms of the normalized parameters: elevation of the sheet above the cathode, $\overline{\xi}$; external resistance, $\overline{R}$; and charge density of the sheet, $\overline{\sigma}$. The derivation given in Supplemental Material leads to
\begin{equation}
 \ddot{\overline{\xi}}+\overline{R}\overline{\sigma}\dot{\overline{\xi}}-\overline{\sigma}\overline{\xi}=1-\frac{\overline{\sigma}}{2} \ ,%
\label{eq:Diffeq}
\end{equation}
with initial conditions $\dot{\overline{\xi}}=0$ and $\overline{\xi}=0$. The solution is
\begin{equation}
 \overline{\xi}(\overline t)=\left(\frac{s_2}{s_1-s_2}e^{(s_1\overline{t})}-\frac{s_1}{s_1-s_2}e^{(s_2\overline{t})}+1\right)\frac{\overline{\sigma}-2}{2\overline{\sigma}} \ ,%
\label{eq:position}
\end{equation}
where 
$s_{1,2} = \left(-\overline{\sigma}\overline{R}\pm\sqrt{(\overline{\sigma}\overline{R})^2+4\overline{\sigma}}\right)/2$. 
The normalizing factors are: $D$ for length, $\sqrt{\frac{mD^2}{qV_0}}$ for time, $\frac{V_0\varepsilon_0}{D}$ for charge density, $\frac{D^2}{L^2\varepsilon_0}\sqrt{\frac{m}{qV_0}}$ for resistance, $\sqrt{\frac{q}{m}}\frac{L^2V_0^{3/2}\varepsilon_0}{D^2}$ for current, and $V_0$ for potential. Using these factors for the parameters of the system shown in Figure~\ref{fig:VoltageIncrease} we obtain, $\overline{R}=57.4$ and $\overline{\sigma}=0.030 \pm 0.001$ (The parameter $\sigma = \frac{qN_e}{L^2}$ with $N_e = 3540 \pm 80$ is obtained from the simulation). The orange curves in Figure~\ref{fig:VoltageIncrease} show how the single sheet model, using these parameters, matches the simulation data. The current and anode potential match quite well for the first part of the oscillation, but deviate from the simulation toward the latter part. This is due to the simple model assuming a sheet of charge, whereas the simulation more accurately shows a clump of charge with finite width such that the electrons are absorbed at the anode over a small period of time rather than instantaneously. This type of broadening of the electron bunch from cathode to anode can be seen in Figure~\ref{fig:VoltageIncrease} (c) where the red dots show the number of electrons absorbed per time step at the anode.

The period of the oscillations can be obtained by solving the equation $\overline{\xi}(\overline{t})=1$. We notice that $s_2<0, \ s_1>0$, and $\mid s_2\mid>s_1$, which means that the second exponential function in Eq.\ (\ref{eq:position}) decreases fast in time and can be discarded.  Assuming a small charge density, such that $\overline{\sigma}\ll 1$, and a large resistance, such that $\overline{\sigma}\overline{R}^2\gg 1$, like in the example discussed above, we can perform a linear expansion of the factors involved and we obtain a simple formula for the period, 
\begin{equation}
\overline{T}=\overline{\sigma}\overline{R}+\frac{1}{\overline{\sigma}\overline{R}}
\hspace{5mm} {\rm or} \hspace{5mm} 
T=\frac{QR}{V_0}+\frac{m}{q}\frac{D^2}{QR} \ ,
\end{equation}
in normalized and physical units, respectively. For the example above this formula gives $\overline{T}=2.30$ or $T=0.11$ ps close to what is seen in Fig.\ \ref{fig:VoltageIncrease} where $T=0.123\pm0.002$ ps. The first term is the discharging time $RC$ of a capacitor with $C=Q/V_0$, while the second term is related to the acceleration of electrons within the diode via the $q/m$ ratio. 
Note that this equation is more descriptive than predictive, as the total charge, $Q$, must be obtained from the simulation.

We have demonstrated how persistent oscillations in current can form spontaneously in a nanodiode with DC voltage applied and an external resistor connected in series. It is apparent that the instability is primarily due to beam loading, because of the resistor, having a strong effect on field emission at the cathode. The oscillation is not contingent on the system being shocked, but can develop even as the applied potential rises slowly compared to the time that it takes a single electron to transit the diode gap. A simple analytic model captures the basic behavior of the instability once it sets in. The oscillations are due to transit time effects in the sense that the current, and subsequent beam loading, increases with time as an electron bunch is accelerated through the gap, and does not decrease until the electrons are absorbed at the anode. Figure S3 of Supplemental Material shows the effects of increasing the applied potential, gap spacing, and external resistance.

Though the instability is due to beam loading of the circuit because of the external resistor, it does not stem from a simple relationship between the magnitude of the external resistor and the impedance of the diode, but is dependent on other parameters such as the cathode work function. Our simulation model is perfectly planar, with a rather high voltage needed to extract current. A real system will typically have emission from points with localized field enhancement. In Figure S4 of Supplemental Material we show an example of oscillatory behavior in a diode with emission from a hyperbolic tip, which is a model of a pin-to-plate configuration \cite{Torfason_2016}. The oscillation is less prominent than in the planar example due to the much smaller current, but the mechanism is the same. 

We emphasize that our oscillations are produced solely by electrons, in a pure vacuum environment, in contrast with the case of gas discharge \cite{Levko_2020,Chen_2024}.  Also, space-charge effects are not essential, as in the nanovircator model \cite{Frolov_2015,Kurkin_2016}, where bifurcation of the nonlinear dynamics into regular and chaotic regimes occurs \cite{Frolov_2017}. Therefore our system is much simpler, with only a weak effect of the space charge, as shown in Supplementarl Material, Figure S4 for the tip geometry, and even weaker in a planar configuration. 
Inductive effects may play a role, especially at the high frequency involved. However, a detailed analysis of the role of the different circuit parameters is beyond the scope of this present paper. 

Additionally, our predicted frequency domain is also different from previous works, between 1-10 THz, and tunable via parameters such as the external resistor and gap spacing, whereas in space-charge-based oscillations it typically ranges between 0.1-1 THz.

The mechanism described in this paper, being reliant on field emission with moderate applied voltage, suggests that it might be amenable to construction of a relatively compact THz source of simple design and good temporal control that could be of use in generating THz radiation for the many potential applications of that frequency range.\\

This work was supported by the Air Force Office of Scientific Research under Award no. FA8655-23-1-7003.

\newpage

\renewcommand{\thefigure}{S\arabic{figure}}
\setcounter{figure}{0}

\textbf{\large{Supplemental Material}}\\

\textbf{1. Derivation of Eq.~(5)}\\

Consider a single sheet of charge density $\sigma = \frac{Q}{L^2}$ of infinite extent. The elevation of the sheet above the cathode is denoted by $\xi$. $E_1$ denotes the field strength in the region between the sheet and the anode, and $E_2$ the strength of the field in the region between the cathode and the sheet. With $D$ denoting the gap spacing of the planar diode, $V_a$ the anode potential, $R$ the external resistor, and $V_0$ the potential of the voltage source we obtain from the potential: $E_1(D-\xi) + E_2\xi = V_a$, and from Gauss' law $E_1 - E_2 = \sigma/\varepsilon_0$. The acceleration of the sheet, $\ddot{\xi} = q(E_1 + E_2)/2m$, is obtained from the average force acting on either side of the sheet (half of the mass being accelerated by the leading field and half of the mass accelerated by the trailing field). The Ramo-Shockley current, $I = \frac{Q}{D}\dot{\xi}$ is used to calculate potential drop over the resistor and and yield the anode potential $V_a = V_0 - RI$. From this one obtains the following equation of motion:

\begin{equation}
 \ddot{\xi}+\frac{qL^2}{mD^2}R\sigma\dot{\xi}-\frac{q}{mD\varepsilon_0}\sigma\xi=\frac{qV_0}{mD}-\frac{q\sigma}{2m\varepsilon_0} \ .%
\nonumber
\end{equation}
This equation can be normalized using the following factors: $D$ for length, $\sqrt{\frac{mD^2}{qV_0}}$ for time, $\frac{V_0\varepsilon_0}{D}$ for charge density, $\frac{D^2}{L^2\varepsilon_0}\sqrt{\frac{m}{qV_0}}$ for resistance, $\sqrt{\frac{q}{m}}\frac{L^2V_0^{3/2}\varepsilon_0}{D^2}$ for current, and $V_0$ for potential. The normalized form of the equation is:
\begin{equation}
 \ddot{\overline{\xi}}+\overline{R}\overline{\sigma}\dot{\overline{\xi}}-\overline{\sigma}\overline{\xi}=1-\frac{\overline{\sigma}}{2} \ .%
\nonumber
\end{equation}

A comparison was made between system behavior with and without space charge effects on electron emission. 
The simulation used parameters of $V_0{_m} = 1700\,\mathrm{V}$, $D = 1000\,\mathrm{nm}$, $L = 1000\,\mathrm{nm}$, 
$R = 240\,\mathrm{k}\Omega$, $\phi = 2.0\,\mathrm{eV}$, and $t_r = 0\,\mathrm{ps}$. 
The results indicate that the influence of space charge effects on emission is negligible under these conditions.

\newpage

\textbf{2. Additional figures}\\


\begin{figure}[H]
  \centering
  \includegraphics[width=\columnwidth]{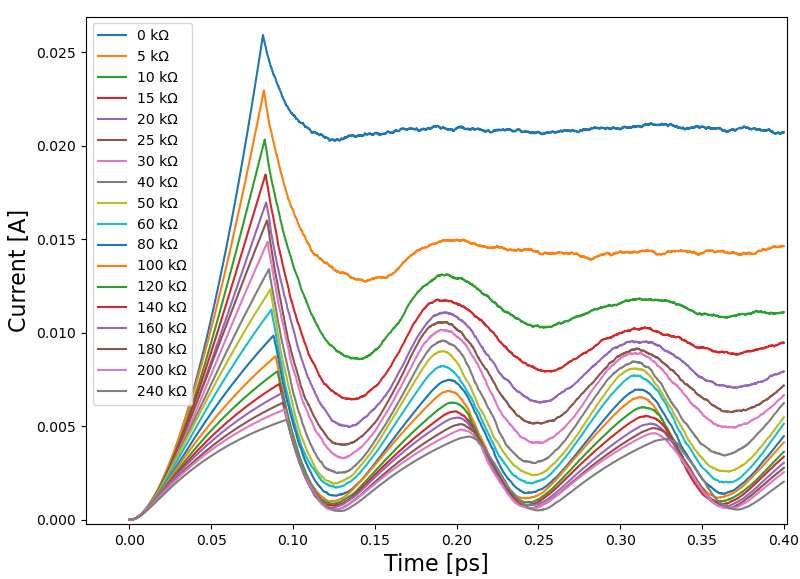}
  \caption{System behavior for increasing resistance values. Simulation parameters: $V_{0m} = 1700\,\mathrm{V}$, $D = 1000\,\mathrm{nm}$, $L = 1000\,\mathrm{nm}$, $\phi = 2.0\,\mathrm{eV}$, and $t_r = 0\,\mathrm{ps}$. A minimum resistance of approximately $60\,\mathrm{k}\Omega$ is required to sustain oscillations under these conditions.}
  \label{fig:different_resistance}
\end{figure}

\begin{figure}[H]
  \centering
  \includegraphics[width=\columnwidth]{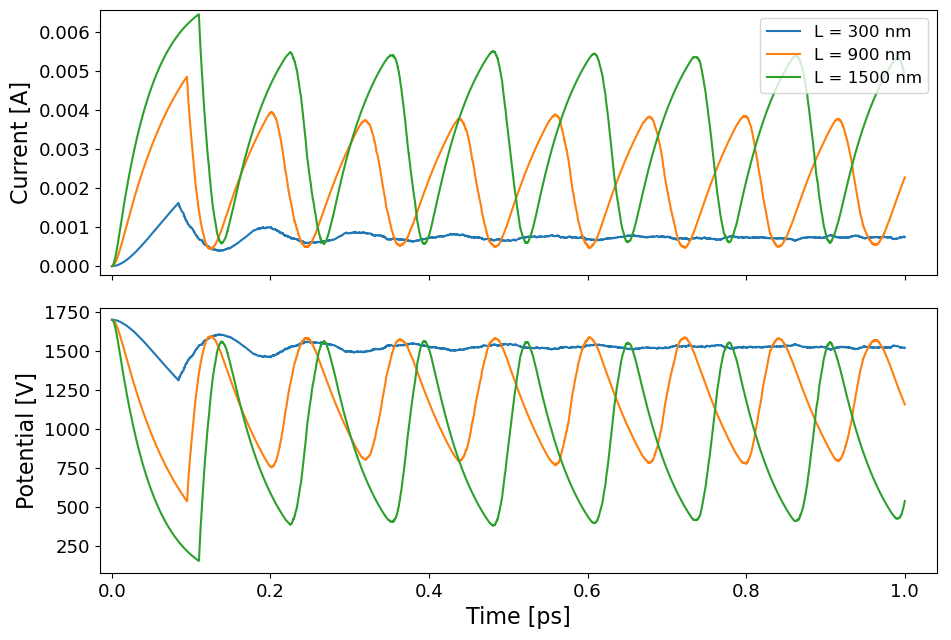}
  \caption{System behavior for different emitter side lengths. 
    (Blue) $L = 300\,\mathrm{nm}$, (orange) $L = 900\,\mathrm{nm}$, (green) $L = 1500\,\mathrm{nm}$. 
    Simulation parameters: $V_0{_m} = 1700\,\mathrm{V}$, $D = 1000\,\mathrm{nm}$, $R = 240\,\mathrm{k}\Omega$, 
    $\phi = 2.0\,\mathrm{eV}$, and $t_r = 0\,\mathrm{ps}$. 
    Increasing $L$ results in longer oscillation periods due to a reduced average voltage, 
    which slows electron transit. A minimum emitter side length is required for sustained oscillations; in this case, $L \geq 500\,\mathrm{nm}$.
  }
  \label{fig:L_differant}
\end{figure}

\begin{figure}[H]
  \centering
  \includegraphics[width=\columnwidth]{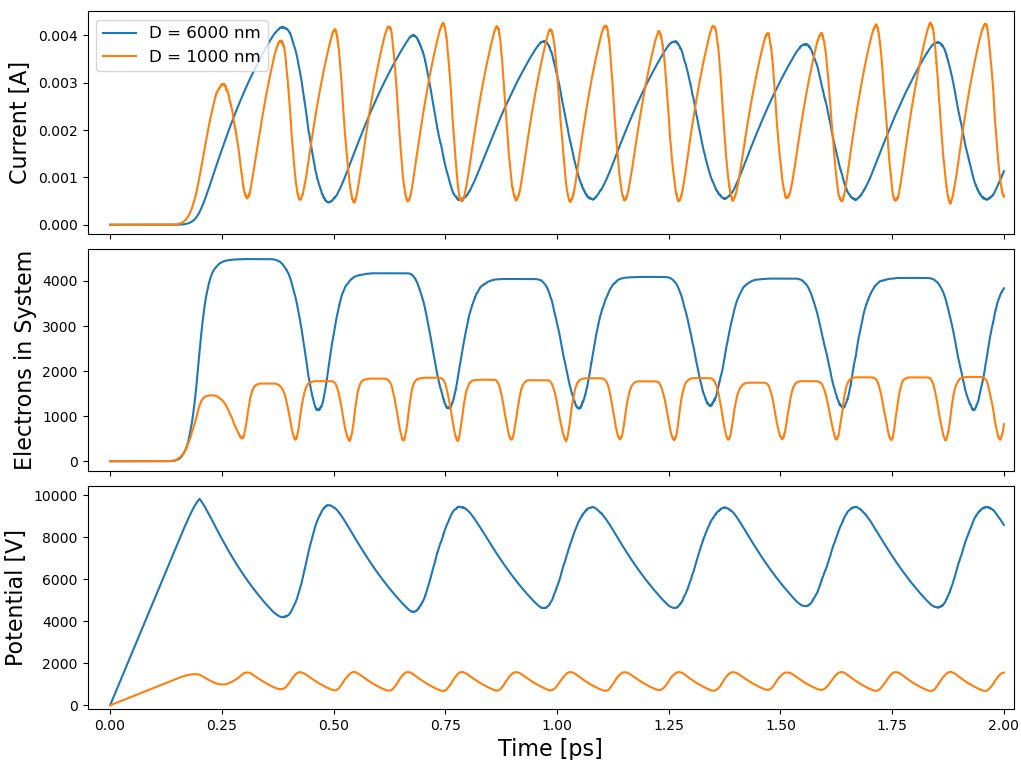}
  \caption{Comparison between $D = 1000\,\mathrm{nm}$ (orange) and $D = 6000\,\mathrm{nm}$ (blue) under identical surface electric field density conditions 
    (i.e., $\frac{V_0}{D}$ is the same for both cases). Both simulations use 
    $L = 1000\,\mathrm{nm}$, $\phi = 2.0\,\mathrm{eV}$, and $t_r = 0.2\,\mathrm{ps}$. 
    For the orange curve, $V_0{_m}= 1.7\,\mathrm{kV}$, $D = 1000\,\mathrm{nm}$, and $R = 240\,\mathrm{k}\Omega$; 
    for the blue curve, $V_0{_m} = 10.2\,\mathrm{kV}$, $D = 6000\,\mathrm{nm}$, and $R = 1440\,\mathrm{k}\Omega$. 
    The sixfold increase in external resistance was chosen to approximately scale with the diode's impedance $Z_0$ as the applied potential and gap spacing were increased sixfold. 
    The results indicate that oscillations persist even at a gap width of $6\,\mu\mathrm{m}$, confirming their presence at the micrometer scale. Note that the total charge in the gap scales as approximately $\sqrt{6}$ in this case.
  }
  \label{fig:high_gap_distance_3000}
\end{figure}
\FloatBarrier

\begin{figure}[H]
  \centering
  \includegraphics[width=\columnwidth]{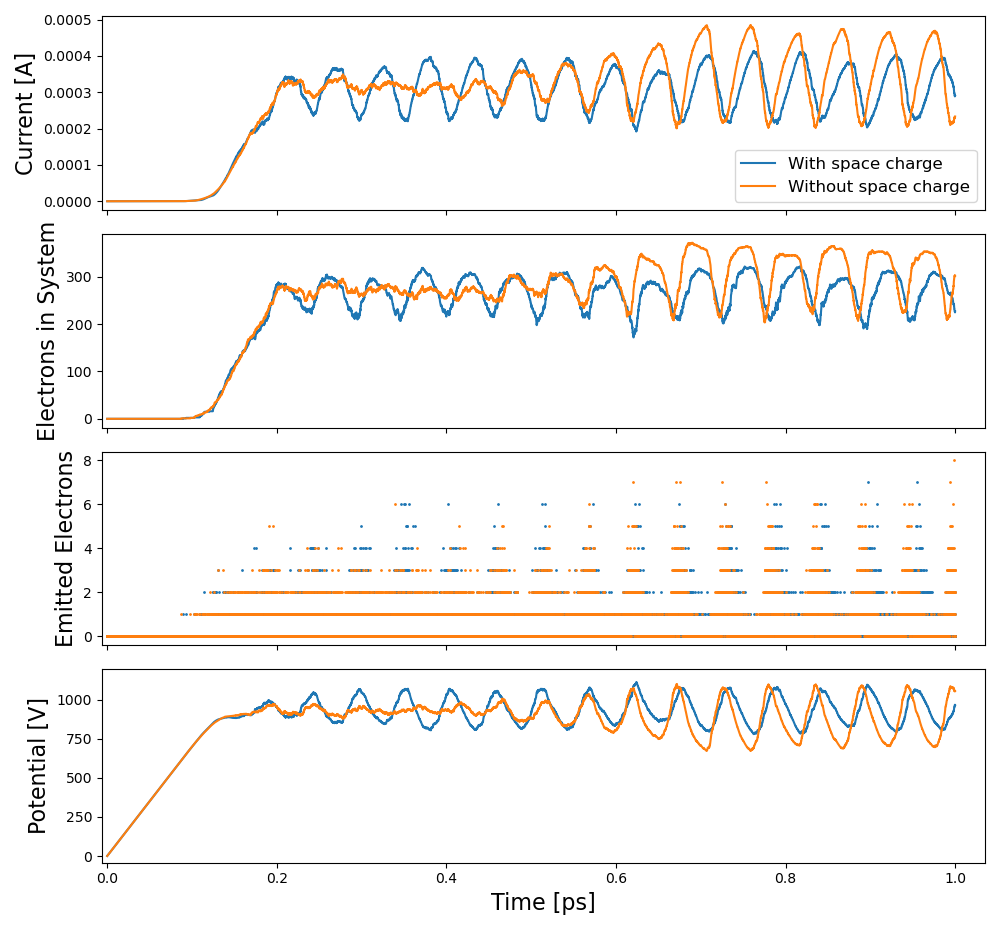}
  \caption{Emission from a hyperbolic tip. $V_0{_m} = 1.4\,\mathrm{kV}$, $R = 1500\,\mathrm{k}\Omega$, $\phi = 2.0\,\mathrm{eV}$, and $t_r = 0.2\,\mathrm{ps}$. The geometry of the tip is shown in Fig.~\ref{fig:tip_explain}, with $r = 250\,\mathrm{nm}$, $h = 500\,\mathrm{nm}$, and $d = 1000\,\mathrm{nm}$. The orange curve includes no space-charge effects, while the blue curve includes space-charge. For the latter, the anode potential remains below $1.1\,\mathrm{kV}$, even though the applied potential is $V_0{_m} = 1.4\,\mathrm{kV}$. Space-charge has an effect on the results, but is not the main driver of the instability.}
  \label{fig:tip_emission}
\end{figure}

\begin{figure}[H]
  \centering
  \pgfdeclarelayer{back}
  \pgfsetlayers{back,main}

  \tikzsetnextfilename{Coords}
  \begin{tikzpicture}
    \def\ximax{2.0}
    \def\etatip{-0.975}
    \def\afoci{1.75}
    \def\yshift{0.0}
    
    \pgfmathsetmacro{\Rbx}{\afoci * sqrt((\ximax)^2 - 1) * sqrt(1 - (\etatip)^2)}
    \pgfmathsetmacro{\Rby}{\afoci * \ximax * \etatip - \yshift}
    \coordinate (R_base) at ($(\Rbx, \Rby)$);
    
    \pgfmathparse{\afoci * \ximax * \etatip - \yshift}
    \coordinate (tipbot) at ($(0.0, \pgfmathresult)$);
    
    \pgfmathsetmacro{\tiptopy}{\afoci * 1.0 * \etatip - \yshift}
    \coordinate (tiptop) at ($(0.0, \tiptopy)$);
    
    \draw[thick] (-3.0, 0.0) node[above, right, yshift=+0.2cm] {} -- (3.0, 0.0) node[above, left, yshift=+0.2cm] {$V_0$};
    \draw[thick, samples=40, join=round] 
      plot[domain=\ximax:1] ({\afoci * sqrt((\x)^2 - 1) * sqrt(1 - (\etatip)^2)}, {\afoci * \x * \etatip - \yshift})
      -- plot[domain=1:\ximax] ({-\afoci * sqrt((\x)^2 - 1) * sqrt(1 - (\etatip)^2)}, {\afoci * \x * \etatip - \yshift})
      node[left, align=center] {\(V = 0\)};
    
    \draw[thick, decorate, decoration={brace,amplitude=10pt}, xshift=-4pt, yshift=0pt]
      (-0.5, \tiptopy) -- (-0.5, -0.075)
      node [black, midway, xshift=-0.6cm] {\(d\)};
    
    \begin{pgfonlayer}{back}
      \begin{scope}[very thick, decoration={markings, mark=at position 0.5 with {\arrow{<}}}] 
        \foreach \i in {1.00, 1.025, 1.15, 1.35, 1.55, 1.75, 1.95}
        {
          \draw[postaction={decorate}, gray, thin, samples=50, join=round, dashed]
            plot[domain=\etatip:0.0] ({\afoci * sqrt(\i^2 - 1.0) * sqrt(1.0 - (\x)^2)*(+1)}, {\afoci * \i * \x - \yshift});
          \draw[postaction={decorate}, gray, thin, samples=50, join=round, dashed]
            plot[domain=\etatip:0.0] ({\afoci * sqrt(\i^2 - 1.0) * sqrt(1.0 - (\x)^2)*(-1)}, {\afoci * \i * \x - \yshift});
        }
      \end{scope}
    \end{pgfonlayer}

    \draw[dotted, join=round] (tiptop) -- (tipbot)
      node[midway, left] {\(h\)}
      -- (R_base) node[midway, below] {\(r\)};
  \end{tikzpicture}

  \caption{Shape of the hyperbolic tip used in simulations presented in Fig.~\ref{fig:tip_emission} \cite{Torfason_2016}.}
  \label{fig:tip_explain}
\end{figure}
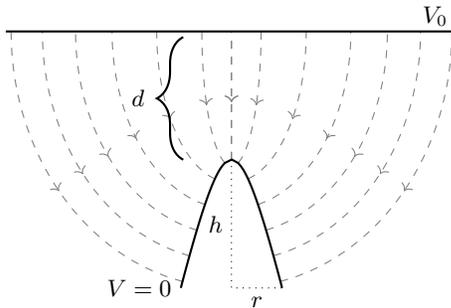
\FloatBarrier

\bibliography{references}

\end{document}